\documentclass[11pt]{article} 
\usepackage{cite}
\usepackage{epsfig}
\usepackage{amsmath,amssymb}

\def\shiftdown#1{#1\llap{\lower.04ex\hbox{#1}}}

\begin{document}
\title{Probing of nucleon mesonic structure by means of\\
quasi-elastic knock-out processes like $p+e\to\ e^{\prime}+\pi^++n$}%
\author{V.G. Neudatchin, I.T. Obukhovsky\footnote{The talk at 
International Symposium MENU2001, The George Washington Univ, 26-31 July.}%
,  and N.P. Yudin\\ 
 \and Institute of Nuclear Physics, Moscow State University\\
Moscow 119899, Russia}
\maketitle
\begin{abstract}
The momentum distributions $\overline{|\Psi^{n\pi^+}_p({\bf
 k})|^2}$ and $\overline{|\Psi^{n\rho^+}_p({\bf k})|^2}$ of pions
and $\rho$-mesons in the nucleon are extracted from experiment.
Perspectives of the quark microscopic theory of mesonic cloud are
outlined.
\end{abstract}
%%%%%%%%%%%%%%%%%%%%%%%%%%%%%%%%%%%%%%%%%%%%%%%%%%%%%%%%%%%%%%%%%%%%%%%%%%%%%%%%%%
\section{Introduction}

It is well known that the theoretical description of the pion
photoproduction on nucleon encounters the complicated set of
problems: a number and a type of used diagrams, their gauge
invariance, necessity of inclusion of form factors, taking into
account off-shell effects, and so on. But the situation becomes
significantly simpler for virtual photons $\gamma^*$ in the
electroproduction process $p+\gamma^*\to n+\pi^+$ at sufficiently
high values of the momentum square transfer $Q^2=-q^2\gtrsim 1-2
GeV^2/c^2$. In this case the diagram with the pion pole in the
$t$-channel (the "pion-in-flight" diagram) becomes dominant under
the standard conditions of quasielastic knock-out process: $|{\bf
q}|\gg |{\bf k}|$, $q_0\gg |k_0|$, $k_0 = M_N-E_N(-{\bf k})$ (or
$W-M_R\gg m_{\pi}$), where \{$k_0,{\bf k}$\} is the 4-momentum of
virtual pion in the nucleon and $W^2=(p_{R}+p_{\pi^{\prime}})^2$
is the mass of final hadron state $R+\pi$ ($R=N,\,\,
\Delta,\,\,N^*$, etc.). For the first time the dominance of
pion-in-flight diagram in the longitudinal part $\sigma_L$ of
differential cross section (integrated over the asimuthal angles)
$d^3\sigma(ep\to
e^{\prime}n\pi^+)/dQ^2dW^2dt=2\pi\Gamma\,(\epsilon\,
d\sigma_L/dt+d\sigma_T/dt), \quad t\approx -{\bf   k^2}$, was
pointed in the paper~\cite{b}. Later in Ref.~\cite{s} the analysis
of the momentum distribution of pions in the nucleon was given in
the frame of light cone dynamics. In our works~\cite{nsy,n} the
discussion of the knock-out process is given in the laboratory
system and just in this frame we find the momentum distribution
(MD) of pions (starting from experimental data \cite{b} on
$\sigma_L$). The similar situation is also for the transverse
virtual photon $\gamma^*_T$ (really the knockout of a vector meson
with simultaneous transformation of $\rho  $ into the pion). It
can provide a valuable information on the properties of vector
mesons in nucleon~\cite{nsy}. These facts open a new possibility
for the direct investigation of the meson structure of nucleon.

%%%%%%%%%%%%%%%%%%%%%%%%%%%%%%%%%%%%%%%%%%%%%%%%%%%%%%%%%%%%%%%%%%%%%%%%%%%%%%%%%%
\section{Study of the pion and $\rho$-meson momentum distribution
in the nucleon}

Our approach to the problem is very similar to the standard
methods of nuclear physics, where the process of nucleon knockout
has been used for a long time as a mighty tool for investigation
of the momentum distribution of nucleons in nuclei. The cross
section of quasi-elastic pion knockout from nucleon can be written
in the form:
 $$
 \frac{d\sigma_L(\gamma^*p\to n\pi^+)}{d{\bf
 k^2}}=\frac{\alpha\,F^2_{\pi}(Q^2)}{8|{\bf q}^*|W(W^2-M_N^2)}\,
 \overline{\Bigl|\frac{M(p\to n+\pi^+)}{t-m_{\pi}}\Bigl| ^2}\,
 \frac{{\bf q}^2}{Q^2}\,\,4\left[\omega({\bf p}_{\pi^{\prime}})-
 q_0\frac{{\bf p}_{\pi^{\prime}}\cdot{\bf q}}{ {\bf q}^2}
 \right]^2,
 $$
where asterisk symbols are for values written in the center of
mass of the $\gamma^*+p$ collision.  The "wave function" (W.F.) of
a virtual pion in the nucleon is the following k-dependent factor
of $\sigma_L$:
 $$
 \overline{\Bigl|\Psi^{n\pi^+}_p(|{\bf k}|)\Bigl|^2}={\cal N}^{-1}
 \,\overline{\Bigl| \frac{M(p\to
 n+\pi^+)}{k_0-\omega_{\pi}(|{\bf k }|)}\Bigl| ^2},\qquad
 {\cal N}^{-1}= \frac{4\pi}{(2\pi)^32M_N\,2E_N(|{\bf k}|)\,
 2\omega_{\pi}(|{\bf k}|)},
 $$
 which can be related to the matrix element of the pion creation
 operator
 $$
 \langle n+\pi^+|a^{\dag}_{\pi^+}|p\rangle\sim
 \frac{M(p\to n\pi^+)}{t-m_{\pi}}=\frac{{\cal N}^{1/2}}{k_0+
 \omega_{\pi}(|{\bf k}|)}\Psi^{n\pi^+}_p(|{\bf k}|).
 $$
Recall that for the standard $\pi NN$ vertex the transition
amplitude can be written in the form
 $$
 \overline{|M(p\to n\pi^+)|^2}=2{\bf k}^2g^2_{\pi NN}F^2 _{\pi NN}
 ({\bf k}^2),\quad F_{\pi NN}({\bf k}^2)=\Lambda_{\pi}^2/
 (\Lambda_{\pi}^2+{\bf k}^2).
 $$
 As the normalization of all the above factors is fixed the norm of
such W.F. defines a "spectroscopic factor" (S.F.) of the $n+\pi^+$
state in the proton
 $$
 S_p^{n\pi}=\int_0^\infty\overline{\Bigl|\Psi^{n\pi^+}_p(k)\Bigl|^2}
 k^2dk,
 $$
The pion-baryon structure of nucleon can be described in terms of
a set of such spectroscopic factors: $S_p^{\pi n}$, $S_p^{\pi
\Delta}$, $S_p^{\pi N^*}$, etc. When the longitudinal cross
section is factorizable in the form:
 $$
 d\sigma_L(\gamma^*p\to n\pi^+)/d{\bf k}^2\sim
 \overline{\Bigl|\Psi^{n\pi^+}_p(|{\bf
 k}|)\Bigl|^2}\,Q^2F_{\pi}^2(Q^2),
 \quad F_{\pi}(Q^2)=(1+Q^2/0.51)^{-1}
 $$
(e.g. in terms of the pion-in-flight mechanism) the W.F. and S.F.
should be observable values in the coincidence experiments (at
fixed $Q^2$ and $M_R$). Then one can define (in analogy with the
standard definitions in the nuclear cluster physics) the " total
number of pions in the nucleon" as a sum $S_{\pi}=\sum_B
S_{p}^{\pi B}$, where $R\equiv B=N$, $\Delta$, $N^*$, etc. are all
the possible virtual baryons in the nucleon.
\begin{figure}[t ]
\parbox{.7\textwidth}
 {\epsfig{file=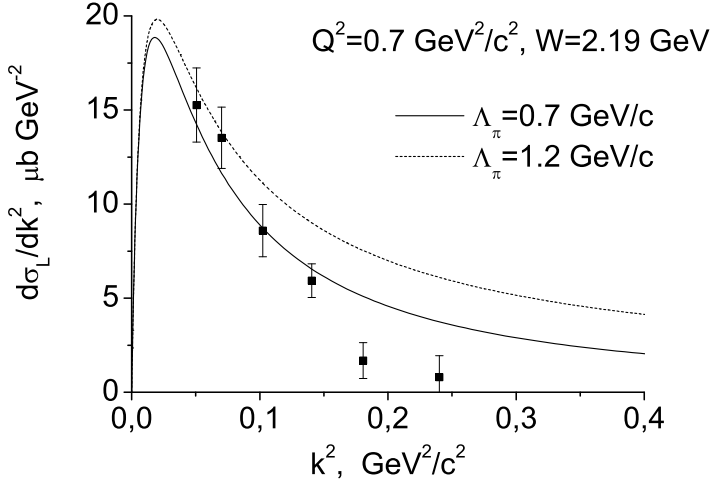,width=0.65\textwidth}
 \caption{\label{f1}Longitudinal cross section at $Q^2=0.7\,
 GeV^2/c^2$. The data are from Ref.~\cite{b}}}
\end{figure}
%-------------------------------------------------------------------
In Figs. 1 and 2 the longitudinal cross sections calculated on the
basis of the above pion-in-flight mechanism (at the value of
$\Lambda_{\pi}= $0.7 and 1.2 GeV) are compared with the old
$p(e,e^{\prime}\pi^+)n$ data~\cite{b} at $Q^2=$0.7 and 3.3
$GeV^2/c^2$.
%--------------------------------------------------------------------
One can see that at both high and intermediate $Q^2$ this
mechanism is in a rough agreement with the data if the cut-off
parameter $\Lambda_{\pi}$ is not so large ($\approx\,$0.7 GeV/c),
but new more exact data on $\sigma_L$ at high $Q^2$ would be
desirable.
%--------------------------------------------------------------------
\begin{figure}[t]
 \parbox{.47\textwidth}
   {\epsfig{file=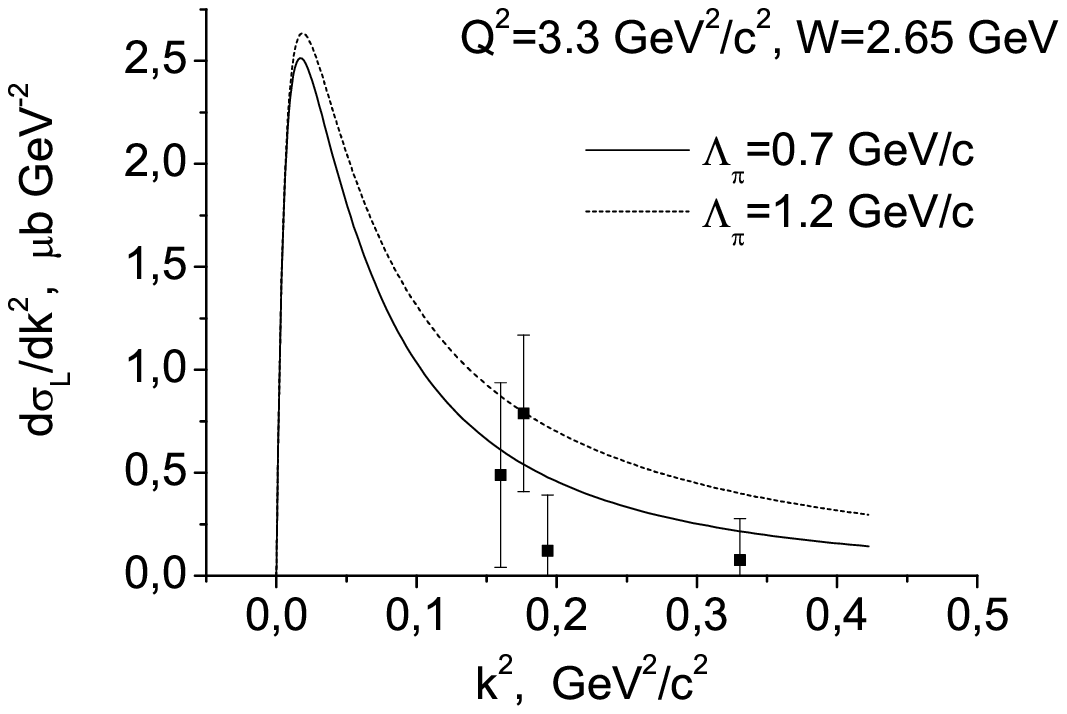,width=0.45\textwidth}
 \caption{\label{f2}Longitudinal cross section at $Q^2=3.3\,
 GeV^2/c^2$.}}
  \hfill
 \parbox{.47\textwidth}
   {\epsfig{file=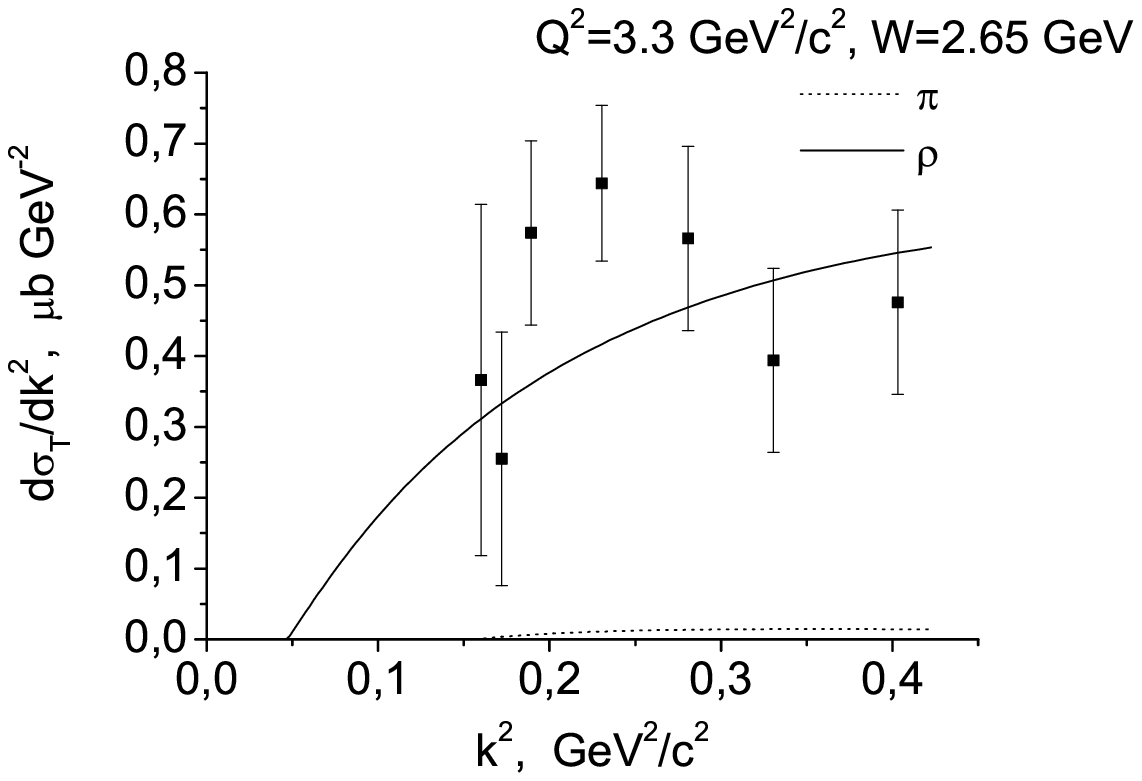,width=0.45\textwidth}
 \caption{\label{f3}Transverse cross section at $Q^2=3.3\,
 GeV^2/c^2$.}}
\end{figure}
%-----------------------------------------------------------------------
\begin{figure}[t]
\parbox{.47\textwidth}
 {\epsfig{file=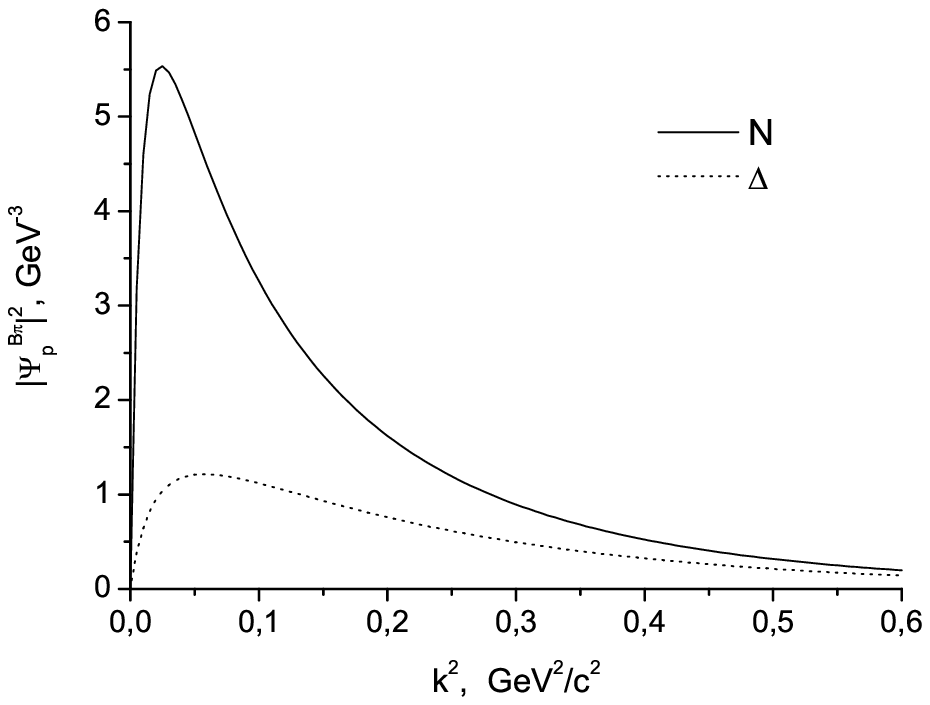,width=0.45\textwidth}
 \caption{\label{f4}Pion momentum distribution in the nucleon
 for $\pi B  $ channels $\pi+N$ and $\pi+\Delta$.}}
 \hfill
 \parbox{.47\textwidth}
   {\epsfig{file=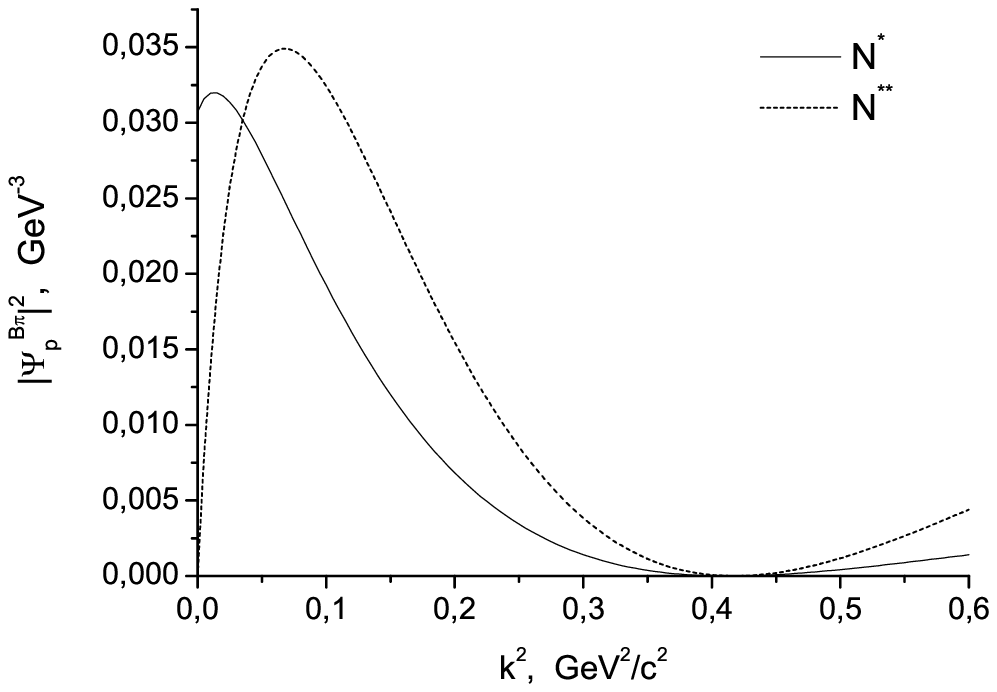,width=0.45 \textwidth}
 \caption{\label{f5}Pion momentum distribution in the nucleon
 for $\pi B$ channels $\pi+N^*$ and $\pi+N^{**}$.}}
\end{figure}

 The pole diagram with the $\rho$ meson as a virtual state in the
nucleon provides the main contribution to the transverse cross
section $\sigma_T(\gamma^*p\to n\pi)$ at high $Q^2\,\gtrsim\,2-3\,
Gev^2/c^2$ (see Fig. 3). However, the spectroscopic factor for the
$n+\rho$ channel $S_p^{n\rho}$ cannot be determined from the
experiment as the data are only available for too low ${\bf k}^2$
as compare with $m_{\rho}^2$.

%%%%%%%%%%%%%%%%%%%%%%%%%%%%%%%%%%%%%%%%%%%%%%%%%%%%%%%%%%%%%%%%%%%%%%%%
\section{\textmd Pion-baryon structure of the nucleon in the $^3P_0$
model}

Non-diagonal transitions $p\to B+\pi$ for the pion-in-flight
mechanism of pion knock out can be considered on the same footing,
i.e. on the basis of the field-theory vertex function, but the
standard expressions
 $$
 \overline{\Bigl|M(p\to B_i+\pi)\Bigl|^2}=f^2_{\pi NB_i}
 \frac{4M_NM_{B_i}}{m_{\pi}^2}2{\bf k}^2F^2_{\pi NB_i}({\bf k}^2)
 $$
depend on too many free parameters ($f_{\pi NB_i}$ and $F^2_{\pi
 NB_i}({\bf k}^2)$ in each channel "$i$") to be informative for
definite predictions of cross sections. The constituent quark
model evaluations of the above free parameters would be more
effective. For example, the microscopic $^3P_0$ model~\cite{f,d}
starts from only one free parameter, the amplitude of the vacuum
$q\bar q$ fluctuation (normalized on the effective $\pi qq$
coupling constant $f_{\pi qq}=\frac{3}{5}f_{\pi NN}$), but predict
all the coupling constants $f_{\pi NB_i}$ and form factors $F_{\pi
 NB_i}$ of the interest starting from the standard
quark-shell-model techniques~\cite{n,o}. Parameters of the
constituent quark model have already been fixed, and are not free
in our approach. All the form factors $F_{\pi NB_i}$ only depend
on two quark-shell-model parameters $b_N$ and $b_{\pi}$ (quark
radii of $N$ and $\pi$) fixed earlier. At the standard values
$b_N=0.6\,fm$ and $b_{\pi}=0.3\,fm$ we have obtained the
predictions for $\pi$-$B_i$ wave functions in the nucleon (see
Figs. 4 and 5), where $B_i=N,\,\Delta,\,N^*=N_{1/2^-}(1535)$ and
 $N^{**}=N_{1/2^+}(1440)$. In the used proper normalization
 of the W.F.'s we are able to determine the S.F.'s for these virtual
states in the nucleon~\cite{o}. This values could  be extracted
from coincidence experiments similar to the above cited~\cite{b}.
Such experiments would be more difficult because of too small
predicted cross sections for $\pi +B_i$ channels, but they will be
very informative for understanding the hadron structure in terms
of quark and meson degrees of freedom.
 
%%%%%%%%%%%%%%%%%%%%%%%%%%%%%%%%%%%%%%%%%%%%%%%%%%%%%%%%%%%%%%%%%%
\section*{Acknowledgments}

The authors gratefully acknowledge the
contribution of Dr. L.L. Sviridova to the work. The work was
supported in part by the Russian Foundation for Basic Research
grant No 00-02-16117.

%%%%%%%%%%%%%%%%%%%%%%%%%%%%%%%%%%%%%%%%%%%%%%%%%%%%%%%%%%%%%%%%%%


\begin{thebibliography}{99}  
\bibitem{b}
     B. Brauel, T. Canzler, D. Cords, et al., Z.Phys. C \textbf{3},
     101 (1979).
\bibitem{s}
     F. Guttner, G. Chanfray, H.P. Pirner, and B. Povh, Nucl.
     Phys. A\textbf{429}, 389 (1984).
\bibitem{nsy}
     V.G. Neudatchin, L.L. Sviridova, and N.P. Yudin, Yadernaya
     Fizika \textbf{62}, 694 (1999).
\bibitem{n}
     V.G. Neudatchin, I.T. Obukhovsky, and N.P. Yudin, {\it
     Relativistic Nuclear Physics and Quantum Chromodynamics},
     Proc. XIV Int. Seminar on High Energy Physics Problems
     (Editors: A.M. Baldin and V.V. Burov), JINR, Dubna (2000),
     vol. 1, p. 55.
\bibitem{f}
     A. LeYaouanc, Ll. Oliver, O. P\`ene, and J.-C. Raynal,
     {\it Hadron Transitions in the Quark Model.} Gordon and
     Breach Sci. Publ., N.Y., London, Paris. 1988.
\bibitem{d}
     F. Cano, P. Gonzalez, S. Noguera, and B. Desplanques,
     Nucl. Phys. \textbf{A343}, 331 (1996).
\bibitem{o}
     I.T. Obukhovsky, The contributed paper of this issue, to be
     published

\end{thebibliography}
\end{document}